\documentclass[journal]{IEEEtran}
\usepackage{amsmath,amsfonts}
\usepackage{algorithmic}
\usepackage{array}
\usepackage[caption=false,font=normalsize,labelfont=sf,textfont=sf]{subfig}
\usepackage[style=ieee]{biblatex}
\usepackage{textcomp}
\usepackage{stfloats}
\usepackage{url}
\usepackage{verbatim}
\usepackage{graphicx}
\usepackage{orcidlink}
\usepackage{csquotes}
\usepackage{multirow}
\usepackage{stfloats}
\usepackage{caption}
\def\BibTeX{{\rm B\kern-.05em{\sc i\kern-.025em b}\kern-.08em
    T\kern-.1667em\lower.7ex\hbox{E}\kern-.125emX}}
\DeclareUnicodeCharacter{0301}{\'{e}}
\usepackage{balance}

\begin{document}
\title{Human Participants in AI Research:\\Ethics and Transparency in Practice}
\author{Kevin R. McKee~\orcidlink{0000-0002-4412-1686}
\thanks{K. R. McKee is with Google DeepMind, London,
United Kingdom.}}

\markboth{IEEE Transactions on Technology and Society}%
{Human Participants in AI Research}

\maketitle

\begin{abstract}
    In recent years, research involving human participants has been critical to advances in artificial intelligence (AI) and machine learning (ML), particularly in the areas of conversational, human-compatible, and cooperative AI. 
    For example, {roughly 9\%} of publications at recent AAAI and NeurIPS conferences indicate the collection of original human data.
    Yet AI and ML researchers lack guidelines for ethical research practices with human participants.
    Fewer than one out of every four of these AAAI and NeurIPS papers confirm independent ethical review, the collection of informed consent, or participant compensation.
    This paper aims to bridge this gap by examining the normative similarities and differences between AI research and related fields that involve human participants.
    Though psychology, human-computer interaction, and other adjacent fields offer historic lessons and helpful insights, AI research presents several distinct considerations---namely, participatory design, crowdsourced dataset development, and an expansive role of corporations---that necessitate a contextual ethics framework.
    To address these concerns, this manuscript outlines a set of guidelines for ethical and transparent practice with human participants in AI and ML research.
    {Overall, this paper seeks to equip technical researchers with practical knowledge for their work, and to position them for further dialogue with social scientists, behavioral researchers, and ethicists.}
\end{abstract}

\begin{IEEEkeywords}
Artificial intelligence, human participants, research ethics.
\end{IEEEkeywords}

\section{Introduction} \label{sec:intro}

\begin{displayquote}
``An overarching and inspiring challenge [...] is to build machines that can cooperate and collaborate seamlessly with humans and can make decisions that are aligned with fluid and complex human values and preferences.''~\cite{littman2021gathering}
\end{displayquote}

\begin{displayquote}
``Development of increasingly sophisticated AI capabilities must go hand-in-hand with increasingly sophisticated human-machine interaction.''~\cite{johnson2019no}
\end{displayquote}

\vspace{1em} \IEEEPARstart{P}{roposed} applications of artificial intelligence (AI) frequently invoke images of collaboration and cooperation between humans and artificial systems. Current research endeavors imagine AI and machine learning (ML) systems as assistants, decision aids, and augmentative tools~\cite{littman2021gathering}. Such proposals increasingly highlight the importance of compatibility and interactivity with humans~\cite{brynjolfsson2022turing,dafoe2020open,irving2019ai,johnson2019no,mckee2023humans}.

The goal of developing human-compatible AI presses the need for research with human participants, aimed at understanding and validating the interactive capabilities of AI. In this class of research, humans---and their interactions with artificial agents---are a primary focal point of behavioral or psychological interest. Similar to experiments in psychology and human-computer interaction (HCI), the aim of these studies is to draw general inferences or make generalizable claims, often beyond the sample at hand. Human-participation projects are regularly published at AI-specific conferences such as Neural Information Processing Systems (NeurIPS; e.g.,~\cite{carroll2019utility,du2020ave}), the International Conference on Machine Learning (ICML; e.g.,~\cite{saha2020measuring}), the International Conference on Learning Representations (ICLR; e.g.,~\cite{shih2020critical}), the annual conference for the Association for the Advancement of Artificial Intelligence (AAAI; e.g.,~\cite{freedman2018adapting}), and the annual conference for Autonomous Agents and Multi-Agent Systems (AAMAS; e.g.,~\cite{mckee2022warmth}).
Participants play an additional critical role in both training and evaluating models at the current frontier of AI research (e.g.,~\cite{glaese2022improving,ouyang2022training,geng2023koala}).

This area of AI research is rapidly growing: across 2021, 2022, and 2023, roughly {9\%} of the papers at AAAI and NeurIPS involved the collection of human data ({Appendix A}). {Other fields rely on robust norms for research with human participants, involving not only adherence to specific ethical practices---such as seeking independent ethical review, collecting informed consent, and compensating participants---but also the responsibility to openly communicate those practices in the research record~\cite{hawkins2023ethical}. As a result, a dual commitment to ethical conduct and transparent reporting has become standard practice across psychology and HCI communities.} In contrast, fewer than one out of every three of these AAAI and NeurIPS papers reported {the implementation} of these practices (Table~\ref{tab:papers_summary}; see also
Appendix~\ref{sec:human_data}).
Guidance for responsible and ethical research practice with human participants from psychology, HCI, and other adjacent fields have evidently not reached AI researchers.

This paper aims to bridge this gap by {synthesizing} guidelines for AI researchers to support ethical and transparent work with human participants. Neighboring disciplines---in particular, psychology, HCI, and the field of research ethics---offer key lessons to be adapted within the particular historical and normative context of AI research (Section~\ref{sec:contextual_concerns}).
Drawing on these insights, this paper proposes four ethical principles to guide AI research with human participants %
(Section~\ref{sec:principles})
and outlines a set of concrete guidelines for ethical and responsible conduct of AI research with human participants, tailored specifically for AI researchers who may not have had exposure to such guidance in the past %
(Section~\ref{sec:guidelines}).
\newcommand{\minitab}[2][l]{\begin{tabular}{#1}#2\end{tabular}}
\begin{table*}[t]
    \begin{center}
    \caption{{Reported Ethical Practices in Recent AI Research Involving Human Participants}} \label{tab:papers_summary}
    \begin{tabular}{ |c|c|c|c| }
    \hline
        \multirow{3}{*}{Publication} & \multirow{3}{*}{\minitab[c]{Proportion of papers reporting\\independent ethical review}} & \multirow{3}{*}{\minitab[c]{Proportion of papers\\reporting consent}} & \multirow{3}{*}{\minitab[c]{Proportion of papers\\reporting compensation}} \\
        & & & \\
        & & & \\
        \hline
        \multirow{2}{*}{NeurIPS (2021--2022)} & \multirow{2}{*}{53.6\%} & \multirow{2}{*}{43.8\%} & \multirow{2}{*}{33.6\%} \\
         & & & \\
        \multirow{2}{*}{AAAI (2021--2023)} & \multirow{2}{*}{9.9\%} & \multirow{2}{*}{2.6\%} & \multirow{2}{*}{19.5\%} \\
         & & & \\
         \hline
        \multirow{2}{*}{\textbf{Weighted average}} & \multirow{2}{*}{\textbf{25.8\%}} & \multirow{2}{*}{\textbf{17.5\%}} & \multirow{2}{*}{\textbf{24.6\%}} \\
         & & & \\
        \hline
    \end{tabular}
    \end{center}
    \caption*{{This table presents an exploratory assessment of the proceedings of two major artificial intelligence (AI) research conferences from 2021 to 2023. The evaluation leveraged an automated approach to analyze the proceedings, annotating the full text of every published paper with a large language model. The annotations noted whether each paper involved the collection of original data from human participants and, if so, whether the paper reported the implementation of three key ethical practices: independent ethical review; the collection of informed consent from participants; and the provision of compensation to participants. Cells indicate the proportion of papers---out of those identified as involving human participants---that report on each ethical practice. The bottom row presents a weighted average over both conferences. Overall, the low levels of these practices reported raise concerns about current ethical norms in AI research with human participants. See
    Appendix~\ref{sec:human_data}
    for further details.}}
\end{table*}

\section{Contextual concerns: Why AI research needs its own guidelines} \label{sec:contextual_concerns}

The quest for ethical guidelines in AI research need not start from scratch. Historical milestones in psychology, HCI, biomedical research, and other adjacent fields provide a number of lessons for the AI community. Chief among them are the Belmont Report and the Common Rule, two historical and regulatory artifacts from the United States that have influenced ethical practices around the world (e.g.,~\cite{national1999national, holm2020belmont, mcdonald2008code}; see also~\cite{capron2008legal} and
Appendix~\ref{sec:what_is_the_historical_context}).
The Belmont Report and the Common Rule offer a particularly helpful starting point for AI research, including with the central challenge of defining a scope for research participation (see
Appendix~\ref{sec:who_are_participants}).

Drawing inspiration from the Common Rule and experimental psychology, we can define human participants as living individuals who provide or contribute information to researchers, particularly in systematic investigations ``designed to develop or contribute to generalizable knowledge'' (45 C.F.R. § 46.102). While research participation is essential for the progress of scientific research, it is not an especially dignifying pursuit for participants themselves---involving little agency or control over their involvement.
Following a series of historic lapses in responsible research conduct, biomedical research, experimental psychology, and other human-centered fields developed explicit guidance for conducting studies with human participants (e.g.,~\cite{apa2017ethical, oates2021bps}; see also
Appendix~\ref{sec:what_is_the_historical_context}).
{Since then, research ethics has evolved considerably, incorporating new domains and challenges (e.g., social media and big data;~\cite{elgesem2002special, schroeder2014big, metcalf2016human}) and cultivating new solutions (e.g., dedicated training programs;~\cite{braunschweiger2007citi}).} Frameworks from these nearby human-centered fields offer a helpful point of embarkation for AI research.

Nonetheless, AI research exists as a distinct scientific discipline. Disciplinary boundaries have traditionally kept AI, psychology, and HCI research relatively separate~\cite{grudin2009ai, yang2020re, kogan2020mapping, urban2021deep}. Even as new developments---particularly the growing interest in conversational AI systems---soften those boundaries, AI scientists face challenges that psychologists and other behavioral scientists typically do not. Three normative and historic trends in the AI field---participatory design, crowdsourced dataset development, and the role of corporations---complicate the issues involved in work with human participants, and create a need for guidelines tailored to AI research.

\subsection{Participatory design}

The particular definition of human participation considered here---inheriting from the Common Rule and from experimental psychology---is distinct from \textit{participatory design}, a school of thought running through the history of technology development. Participatory design refers to research approaches that directly ``involv[e] affected populations in shaping the goals of the overall system''~\cite{kulynych2020participatory} (see also~\cite{birhane2022power, simonsen2013routledge, delgado2023participatory}). The origins of participatory design trace back to mid-20th century movements advocating for greater community involvement in public decision making~\cite{schuler1993participatory}. Early Scandinavian experiments in participatory design are particularly well known; these experiments incorporated union workers in the technological design process, reflecting a deep commitment to democratization~\cite{gregory2003scandinavian}.

``Participation'' in this sense is a common calling card in contemporary AI research~\cite{birhane2022power, sloane2022participation}. For instance, many of the organizations responsible for the development of large language models argue that democratic participation must be a core component of future AI systems~\cite{glaese2022improving, roose2023dario, zaremba2023democratic}. The prominence of such appeals places an onus on AI researchers to contend with the relationship between participatory design and their own work with human participants.

Broadly, participatory design places no particular emphasis on creating generalizable knowledge. It instead underscores the importance of remaking power imbalances~\cite{kulynych2020participatory}. Critiquing the misuse of ``participation'' language in machine learning research, \citeauthor{sloane2022participation}~\cite{sloane2022participation} argue that participatory design can manifest in several modes, each varying in its extent of power redistribution---including ``participation as consultation'' and ``participation as justice.'' \citeauthor{arnstein1969ladder}~\cite{arnstein1969ladder} exemplified this focus on power, arguing that ``participation without redistribution of power is an empty and frustrating process for the powerless.'' In contrast with this focus, neither the Common Rule nor most {subfields of} experimental psychology hold redistribution as a central aim. Participants in the sense of the Common Rule often do not qualify as participants in the framework of participatory design, given the latter's emphasis on co-design and co-direction. Nonetheless, several research subfields demonstrate that these definitions of participation can be compatible. {For example, psychiatry researchers routinely involve patients and engage the public in co-design and research planning, particularly in the UK~\cite{rose2014patient, skorburg2024persons}. On a similar note,} participatory action research~\cite{fals1991action} and participatory HCI~\cite{vines2013configuring} advocate for community involvement in setting experimental research agendas, effectively marrying participatory design with human-participant research. Overall, given its historic and contemporary intertwining with participatory design, AI research must similarly reckon with the role of communities in human-participant work.

\subsection{Crowdsourced dataset development}

\textit{Crowdsourced dataset development}, a ubiquitous activity in AI and ML research, similarly challenges the boundaries of traditional ``participant'' definitions.  Supervised learning algorithms learn to predict outputs through exposure to labeled training data (i.e., example input-output pairs). In early periods of the field, expert, in-house labelers generated the annotations for training data. These experts were oftentimes the researchers themselves, though occasionally included students and professional annotators~\cite{snow2008cheap}. The emergence of online crowdsourcing transformed the annotation process, improving the speed, scalability, cost, and convenience of dataset development. AI and ML researchers now crowdsource most labor for data labeling, annotation, and production (see~\cite{gray2019ghost}). The recruitment of human annotators appears to be growing rapidly at the frontier of AI research, given the amount of data needed to train and fine-tune large language models (e.g.,~\cite{glaese2022improving, ouyang2022training}).

A current debate revolves around whether such crowdsourced workers qualify as research participants---``a gray area that is open to interpretation,'' as noted by \citeauthor{shmueli2021beyond}~\cite{shmueli2021beyond}. For example, the Association for Computational Linguistics indicates that ``human annotators (e.g., crowdworkers)'' are distinct from ``human subjects'' in its ``responsible research'' checklist~\cite{aclndresponsible}. In contrast, \citeauthor{sloane2022participation}~\cite{sloane2022participation} argue that such unrecognized contributions to the production and improvement of ML can be considered ``participation as work,'' and thus impart particular ethical responsibilities upon researchers making use of crowdsourcing. The inclusion of annotators as research participants would have profound practical implications: it would afford them protections and oversight that could help protect their well-being and reduce unnecessary harms. Recent incidents across the field's expansive reliance on crowdsourced data (e.g.,~\cite{hao2023cleaning, gray2019ghost}) present a strong argument to include crowdworkers in the framework of human participation for AI research.

\subsection{The role of corporations}

A final concern for AI research revolves around the relationship between participants and researchers. The fields of psychology and HCI have long contended with the imbalanced power dynamics built into the researcher-participant relationship. Most researchers view the sole responsibility of the participant as responding to the research study as they have been directed. The participant's main affordance is to interact with the study materials, and sometimes also with the research team: few other levers are available to them. Indeed, this relative lack of autonomy often emerges in the language used to refer to participants in scientific writing (e.g., ``subjects'' of a research study, rather than ``participants'') and even in regulatory text (e.g., the Common Rule's passive description of a participant as ``an individual about whom an investigator [...] obtains information''; for additional discussion, see
Appendix~\ref{sec:terminology}).
In contrast, the researcher controls most functions outside of data production, including the provision of tools for recourse, compensation for research participation, and the dissemination of research benefits. Because of this imbalance, researchers can readily cause harms when attention to ethical practice lapses, as demonstrated by a long series of oversights in scientific history~\cite{brandt1978racism, gillespie1989research, flick2016informed}.

In AI work, financial and compute resources have proved crucial for recent progress~\cite{ahmed2020democratization, sutton2019bitter}. These resource requirements make technology companies disproportionately influential in AI research, given the particularly large research budgets and caches of compute that they can bring to bear on algorithmic development~\cite{birhane2022values, luitse2021great, martinez2021research}.
The involvement of companies introduces commercial interests into the research process, complicating the researcher-participant relationship. Commercial interests create potential conflicts with the responsibilities that researchers hold toward participants: research teams may face situations where particular tasks or methods can help further industry goals, but are not in the best interests of participants (cf.~\cite{morin2002managing}). Separately, the resources held by technology companies amplify the power imbalance between participants and researchers. Participants in AI research are likely to encounter especially large disparities in financial, legal, and public-relations resources and knowledge between themselves and the teams administering research. {For example, companies studying interactions with state-of-the-art language models may request that potential participants sign confidentiality agreements in order to participate~\cite{ipo2021impact, resnik2006openness}. Even if these agreements allow for disclosure in cases of potential harm, the complex language of standard commercial contracts may stoke fears of legal repercussions or the loss of study compensation, deterring participants from reporting any legitimate risks they encounter.} These imbalances in resources and knowledge may limit participants' agency and access to modes of recourse, especially relative to individuals participating in research in other fields.

\subsection{Summary}

Overall, these challenges create concerns for AI and ML researchers that are not typically shared by scientists in nearby fields. Psychology, HCI, and related fields can offer initial insights and a shared vocabulary for ethical research practice, but as a separate scientific discipline with distinct concerns, AI research necessitates its own frameworks and guidelines.

\section{Ethical principles for AI research with human participants} \label{sec:principles}

Research ethics is both iterative and contextual. The Belmont Report offered a foundation for the Common Rule, which in turn {initiated a series of successive ethics frameworks among experimental psychologists, social scientists, and big data ethicists (e.g.,~\cite{apa2017ethical, metcalf2016human, buchanan2008internet};} see also~\cite{musschenga2005empirical} and
Appendix~\ref{sec:what_is_the_historical_context}).
AI research can similarly iterate by integrating the normative concerns of participatory design, crowdsourced dataset development, and the role of corporations into traditional research ethics frameworks. Through this process, we can consider four ethical principles---\textit{autonomy}, \textit{beneficence}, \textit{justice}, and \textit{accountability}---as a starting point for ethical AI research with human participants. The first three principles emerge directly from the Belmont Report{,} the Common Rule, {and their direct successors,} though contextual concerns give them a new focus. The role of corporations in AI research elevates the final principle, accountability, as another focal point for ethical practice with human participants. 
These four principles provide justification for the practical guidelines presented in Section~\ref{sec:guidelines}. Together, these principles and guidelines form an extended scaffolding for ethical and responsible practice with human participants.

\subsection{Autonomy}

\begin{displayquote}
``The principle of respect for persons thus divides into two separate moral requirements: the requirement to acknowledge autonomy and the requirement to protect those with diminished autonomy.''~\cite{national1978belmont}
\end{displayquote}

Within the Common Rule framework, the ``respect for persons'' principle communicates the idea that researchers should preserve and protect participants’ autonomy. It places an emphasis on the voluntary nature of participation, motivating practices such as the necessity of informed consent for participants and the upfront discussion of risks and benefits of participation.

In the context of AI research, it may be clearer to refer to this concept as the principle of \textit{autonomy}. AI ethics frameworks frequently identify autonomy as a key value for the governance of AI systems, asserting that deployed systems should promote (rather than restrict) freedom and self-determination~\cite{floridi2022unified, jobin2019global}. These ethics frameworks rarely include explicit discussion of research with human participants. Nonetheless, researchers and developers should extend the principle of autonomy to interactions with human research participants. That is, people should be able to choose how and whether to engage with AI systems, both in the real world \textit{and} in the laboratory. As an example of this principle in practice, scientists should transparently communicate with potential participants about their involvement in AI research and development, allowing them to choose whether or not they wish to take part. After all, many individuals may not wish to support the development of new AI systems. AI research has not always met this bar~\cite{gibney2017google}.

\subsection{Beneficence}

\begin{displayquote}
``Two general rules have been formulated as complementary expressions of [beneficence]: (1) do not harm and (2) maximize possible benefits and minimize possible harms.''~\cite{national1978belmont}
\end{displayquote}

The principle of \textit{beneficence} represents two priorities: first, preserving participant well-being, and second, maximally advancing the interests of the participants. It is sometimes distilled to a simple arithmetic rule: do the benefits resulting from the research outweigh the risks?

In AI research, a growing number of initiatives advocate for an analogous calculus for ``social good'' or ``social benefit''~\cite{tomasev2020ai, cowls2021definition}. In the context of human-participants research, such a calculus entails limiting the exposure of each participant to risks and potential harms. Work on algorithmic fairness, for instance, imposes difficult tradeoffs between group equity and individual privacy. Developing AI systems that treat marginalized groups fairly requires collecting data from vulnerable members of those groups, introducing direct risks to the privacy and safety of those individuals~\cite{tomasev2021fairness}. The prevalence of crowdsourced dataset development for state-of-the-art systems raises a number of similar questions. Under what conditions is it ethical to recruit participants to train models that will replace them~\cite{brynjolfsson2022turing}? What sort of protections should developers offer participants who they task with reviewing disturbing or toxic content, when the outcome of research is an AI system intended for use by a broader population~\cite{hao2023cleaning}?

\subsection{Justice}

\begin{displayquote}
``Who ought to receive the benefits of research and bear its burdens? This is a question of justice, in the sense of ‘fairness in distribution’ or ‘what is deserved.’''~\cite{national1978belmont}
\end{displayquote}

Within the Common Rule framework, the principle of \textit{justice} codifies the idea that both the burden and the benefits of research should be equitably shared among potential participants. Its inclusion in the Belmont Report and the Common Rule reflects scientists' and ethicists' concern over historic injustices against marginalized communities, as well as the possibility that future research will continue such exploitation. The justice principle is especially relevant for research that, by design, involves participants from vulnerable or marginalized populations.

Within AI research, who will provide the labor and cost necessary to cultivate progress, and who will reap the benefits?
In the tradition of participatory design, promoting justice requires understanding the structures of power under which research unfolds.
{For instance, many research teams developing image-recognition and image-generation systems seek training data from racial minorities in order to mitigate algorithmic bias~\cite{merler2019diversity, crawford2021excavating}, yet minority communities suffer when the resulting improvements enable applications like predictive policing~\cite{lee2022police} and artificially feigning racial inclusion~\cite{colon2024exploiting}.
Similar tensions emerge around proposals to use contemporary language models to preserve} and revitalize endangered languages. Many indigenous communities still keenly feel scars and trauma from their historic experiences with colonialism{, and---given the involvement of technology companies---}fear that advances in artificial intelligence will be used to exploit their members~\cite{lewis2020indigenous}.
Such cases call for centering the voices, perspectives, and experiences of marginalized groups, in order to properly promote justice.

\subsection{Accountability}

\begin{displayquote}
``To be accountable for [research], researchers must be able to describe in a way sufficient for the social situation at hand how any perceived [research] problems are anomalous, correctable, or in fact not problematic at all---they must be ‘answerable’ for their [research].''~\cite{mayernik2017open}
\end{displayquote}

The principle of \textit{accountability} reflects the responsibility of investigators to explain research processes and outcomes to research participants, to fellow researchers, to academic and public institutions, and to community stakeholders. Such explanations include discussion of the validity, benefits, and harms of research. In this respect, the principle of accountability also encompasses an openness to scrutiny concerning potential issues with research processes and outcomes. As \citeauthor{makel2017toward}~\cite{makel2017toward} explain, ``at its foundational level, the heart of science is that its methods allow for others to believe its results.'' 
\citeauthor{klein2018practical}~\cite{klein2018practical} similarly argue that the cumulative and self-corrective properties of science operate best ``when the scientific community is able to access and examine the key products of research.'' {Overall, accountability encompasses the broad responsibility of researchers to justify and hold responsibility for the potential consequences of their work (cf.~\cite{mayernik2017open}).}

Those involved in psychology and HCI research have issued repeated calls for accountability and social responsibility in scientists’ relationship with the public~\cite{bazelon1982veils, deleon1988public, grimpe2014towards}. Nonetheless, the Belmont framework does not make accountability focal as a principle. The AI community, in contrast, widely recognizes that accountability is a lodestone value for algorithmic deployment~\cite{diakopoulos2016accountability, floridi2022unified, jobin2019global}.
For instance, many advocate for the establishment of clear lines of responsibility for the outputs of deployed systems and identifiable mechanisms for recourse and redress when deployed systems cause people harm.
Accountability is similarly essential for AI work with human participants, on account of the conflicting interests introduced by corporations and competitive interests.
``Ghost work'', the hidden and exploitative labor of data annotation and other essential tasks for AI development~\cite{gray2019ghost}, exemplifies the harmful practices that can proliferate when algorithmic development and evaluation neglect accountability.

\section{Principles in practice: Guidelines for AI research with human participants} \label{sec:guidelines}

Autonomy, beneficence, justice, and accountability provide a crucial framework for navigating the complex landscape of AI research with human participants. However, relying on this framework alone would be like navigating with only a compass: it offers a general sense of direction, but fails to provide a specific route for travel. To best ensure ethical research conduct, researchers still need a method for translating these abstract ideals into actionable steps within specific research contexts~\cite{mittelstadt2019principles}. In other scientific fields, guidelines act as just such a map, helping researchers translate ethical principles into concrete actions.
This section draws from the four principles, in addition to sources from psychological and behavioral research (e.g.,~\cite{apa2017ethical, oates2021bps}), and presents a set of guidelines tailored for AI research with human participants. This guidance can be separated into three groups: guidelines for before the study, during the study, and after the study. To emphasize the theoretical motivation for this guidance, each guideline identifies its corresponding principle or principles in italics.

\subsection{Before the study: Independent ethics review}

At the beginning of the research process, the research team takes several important steps that lay the conceptual foundation for ethical interactions with participants. All four principles motivate this ethics groundwork.

\textbf{Undergo independent ethical review.} Researchers should subject their proposed research studies to independent ethical review before starting any data collection or processing. Independent ethical review should be undertaken by a group with a mixture of scientific experts, community representatives, and at least some members knowledgeable of research ethics. Group members should neither be affiliated with the research (e.g., a member of the research team) nor possess conflicting interests with regard to research outcomes (e.g., a financial beneficiary of any research outcomes). {Given the substantial time and effort required to carry out their responsibilities, group members should be remunerated---both to support the long-term sustainability of their work and to recognize the pivotal role of ethical review~\cite{catania2008survey}.} Many countries designate official bodies to conduct independent ethical review for research with human participants, such as institutional review boards (IRBs) in the United States and research ethics committees in the United Kingdom and the World Health Organization. In the United States, many argue that IRBs have become overly bureaucratized, focusing more on procedural concerns than substantive ethical questions~\cite{heimer2010bureaucratic, white2007institutional}. In principle, the purpose of independent review is not to subject data collection to excessive bureaucratic work and yellow tape, but rather to think earnestly about whether a proposal advances ethical research principles. As a result, various organizations have developed a new class of analogous ethics bodies outside the IRB framework, including the Ethics Review Program at Microsoft Research {(established in 2013)}~\cite{microsoftndmicrosoft}, the Institutional Review Committee at Google DeepMind {(established in 2019)}~\cite{gdmndhuman}, and the Ethics and Society Review board at Stanford University {(established in 2020)}~\cite{bernstein2021esr}.
Independent ethical review is imperative for AI researchers undertaking work with human participants, supporting each of the four principles: \textit{autonomy}, \textit{beneficence}, \textit{justice}, and \textit{accountability}.

\textbf{Solicit peer feedback.} To supplement formal review processes, researchers should consider seeking out informal feedback on their research proposals from their peers. Peers are often well positioned to evaluate and offer comment on proposed studies, both with their disciplinary expertise and from any ethical concerns they have. Relative to reviews by institutional committees, peer feedback may more effectively assess ``technical'' aspects of research, such as the importance of a research question or the validity of a proposed study design. In addition, institutional committees may not always be aware of the capabilities and behavior of state-of-the-art models and agents. Inappropriate design choices may undermine results and inferences, reducing the benefits of research (as per the principle of beneficence). Given their on-the-ground perspective, peer feedback on ethical considerations may also serve as a useful supplement to formal review. This also offers peers themselves practice with ethical design. Ethical deliberation need not be restricted to institutional review, after all: researchers should have the chance to deliberate on potential dilemmas with their colleagues~\cite{hilppo2019theorizing}. In a decentralized manner, these peer channels can help encourage the development of an ethical community among AI researchers. As with the guideline of seeking formal ethical review, this guideline pertains to each of the four principles: \textit{autonomy}, \textit{beneficence}, \textit{justice}, and \textit{accountability}.

\textbf{Avoid deception, if possible.} Experiments have long made use of deception; for nearly just as long, psychologists and other social scientists have debated the acceptability of deception as an experimental practice~\cite{herrera1997historical, ortmann2002costs}. These questions are especially critical for AI research. Should researchers alert or disclose to participants that they are interacting with AI systems? More broadly, should researchers inform participants of the sorts of systems they may be helping to develop? Concealing the nature or goals of AI research from participants inherently conflicts with the principles of \textit{autonomy} and \textit{beneficence}. Researchers should have a cogent justification for deceiving participants (e.g., investigating the effects of ``dishonest anthropomorphism'';~\cite{leong2019robot}). In studies where deception is justified, researchers should debrief participants as soon as data collection is complete, informing them of the deception and providing corrected information.

\subsection{During the study: Ethical conduct and engagement with participants}

While conducting a study, the researcher should seek not just to carry out their planned research protocol, but also to affirm participant dignity. The following guidelines are tailored to online research platforms, given the current prevalence of online study administration. Nonetheless, they can also be readily adapted to in-person studies or other modes of research. Overall, this set of guidelines is primarily motivated by the principles of \textit{autonomy} and \textit{beneficence}.

\textbf{Collect (genuinely) informed consent.} The concept of informed consent captures two interwoven concerns. First, researchers should provide any individuals they recruit with as much information about the research at hand as possible. Second, recruited individuals should only participate after processing this information and providing voluntary, decisionally capacitated consent. The consent process should communicate information including the general research topic, the general activities involved, risks and benefits of participation, compensation details, participants’ right to withdraw, and contact information in case participants have any questions or concerns. To minimize the difficulty participants have processing this information, investigators should write forms in easily comprehensible language~\cite{perrault2016informed, perrault2018seeking}. Taking the step of collecting informed consent centrally supports the principle of \textit{autonomy}.

\textbf{Limit risks.} Research participation should entail minimal risk of harm for participants. Researchers should not just consider physical or psychological risks. Potential risks from research can also be social, legal, or economic in nature. The ideal standard is no-greater-than-minimal risk. As defined by the Common Rule, no-greater-than-minimal risk means that ``the probability and magnitude of harm or discomfort anticipated in the research are not greater in and of themselves than those encountered in daily life or during the performance of routine physical or psychological examinations or tests'' (45 C.F.R. § 46.102). Researchers should be able to provide a strong justification of any higher level of risk with specific, concrete benefits or needs, as per the principle of beneficence. Human-AI interaction work merits particular attention to risks. For example, when chatting with human participants, conversational models may produce toxic or harmful language or provide professional advice against a participant’s best interests, despite an initial appearance of helpfulness and harmlessness (see also~\cite{weidinger2022taxonomy}). Researchers should proactively take steps to mitigate these potential harms, including implementing safety mechanisms in model design and testing agents rigorously before deploying in studies. Limiting such risks underpins the principle of \textit{beneficence}.

\textbf{Support dignity.} Research participation is not a particularly empowering experience. Researchers should especially strive to support participants’ dignity. Participants provide researchers the privilege and honor of access to their thoughts and perspectives. In turn, researchers should provide opportunities for participants to voice their opinions, thoughts, and reactions to their studies. Often, this is as simple as including an open-ended debriefing question at the end of the study, asking: ``Do you have any other thoughts you’d like to share with us?'' Despite the frustrations that participants experience during studies, many also feel pride and derive personal meaning from their contributions~\cite{fowler2022frustration, gupta2014understanding} (cf.~\cite{allan2019outcomes}). (Researchers genuinely interested in understanding the experience and perspectives of research participants can read posts on open forums where participants discuss their experiences and ask each other for advice; e.g., \url{https://www.reddit.com/r/ProlificAc/}. These posts offer a glimpse of the ups and downs of research work, including the range of joys and frustrations that participants feel on a daily basis.) An especially important way for researchers to support these positive experiences is to inform participants about the purpose of studies to which they contribute (e.g., disclosing the intended uses of any system their data will help to develop). Even seemingly small moments can support dignity, such as taking inclusive approaches to asking participants about their gender identity when collecting demographic data (e.g.,~\cite{fraser2018evaluating}). Efforts to support dignity enshrine the principles of \textit{autonomy}, \textit{beneficence}, and \textit{justice}.

\textbf{Maintain channels for clear and timely communication with participants.} Online data collection rarely goes as planned. Server connectivity issues, unforeseen bugs in user interfaces, misconfigured study conditions, unclear instructions---a long list of problems can disrupt online studies. Participants consistently cite lack of communication channels and lack of researcher responsiveness among their top frustrations with online research~\cite{berg2015income, fowler2022frustration}. In contrast, researchers in on-site laboratories rarely leave participants unsupervised. Typically, a team member remains on hand to answer questions and troubleshoot any technical difficulties. The same standards should apply to online data collection: researchers hold the responsibility to maintain convenient channels (and remain personally present) for clear and timely communication with participants. Providing such channels reinforces the principles of \textit{beneficence} and \textit{accountability}.

\textbf{Ensure fair compensation.} Low compensation is a substantial problem among online recruitment platforms~\cite{semuels2018internet}. In one study, researchers found that the median online participant on a popular research platform earned only \$2 each hour~\cite{hara2018data}. On these platforms, a participant’s compensation is distributed across time spent completing studies, as well as time spent searching for studies and reading instructions. Researchers can adopt several different models for compensation, including payment per completed task or payment per hour~\cite{partnershiponaindresponsible}. Given the possibility of misestimating the time a task takes to complete, an hourly rate offers a stable, predictable financial structure for both participants and research teams. Researchers should strongly consider {going beyond minimum wage and instead use living wage calculations~\cite{globalndwhat} to ensure adequate study compensation.} Researchers can also consider an adaptive system, supplementing compensation for individual participants who are particularly engaged or perform especially well on a task, potentially up to a specified upper bound. During recruitment, researchers should clearly communicate compensation details, including any conditions for study compensation and any lower or upper limits for studies with flexible compensation. During the study, researchers should monitor the duration of each session, offering bonuses for any ``overtime'' participation. Overall, fair compensation promotes the principle of \textit{beneficence}.

\textbf{Ensure ability to withdraw without consequence.} Participants should be able to freely withdraw from a study, even after initially providing their consent for participation. Withdrawal should not carry any adverse consequences for participants; researchers should still compensate participants for the time they spent on these partial sessions (e.g., by pro-rating compensation). Participants may also request to withdraw their research data from the study. Consent forms should explain as clearly as possible the point at which investigators can no longer identify and withdraw a particular participant’s data. This guideline reinforces the principles of \textit{autonomy} and \textit{beneficence}.

\textbf{Avoid coercion and undue influence.} Coercion and undue influence are two interconnected concepts in research ethics. Coercion refers to implicit or explicit threats to cause harm to an individual or to withhold a benefit from them in order to change their behavior (e.g., coercing them to complete a research study). Coercion can be a concern when inviting employees of a company to participate in research studies, in particular when the investigators work at the same company~\cite{resnik2016employees}. This tension is especially relevant for AI research, given the number of industry researchers in the field. Undue influence manifests when researchers offer excessive amounts of compensation when recruiting participants, typically for studies that may conflict with participants’ interests and well-being. Thus, researchers should be especially careful when working on research involving higher levels of risk (e.g., bias or toxicity work with language models). This guideline supports the principles of \textit{autonomy}, \textit{beneficence}, and \textit{justice}.

\subsection{After the study: Transparent reporting}

This final subset of guidelines identifies information that researchers should publicly report about their work, framed as questions that can be answered in the main text or appendices of publications. {In psychology and HCI, such transparent reporting is considered essential for responsible research conduct. Many journals in these fields mandate the disclosure of key ethical practices as a condition of publication.} Existing sources in AI research outline several (but not all) of these details~\cite{aclndresponsible, neuripsndneurips}. Given their focus on encouraging the transparent and public dissemination of information, each of these guidelines directly supports the principle of \textit{accountability}.

\textbf{Report details of independent ethical review.} Did the protocol receive independent ethical review before data collection began? What was the outcome of that review? What is the name of the body that undertook the assessment? If that body assigned the reviewed protocol an identifier, researchers should provide that identifier. If any peers provided feedback on the ethics of the research, this information should also be included.

\textbf{Report collection of informed consent.} Did participants provide informed consent before data collection began? Did any issues arise during the provision of study information or the collection of informed consent? Were participants informed of the purpose of the study, including the connection to AI research and the involvement of any AI systems?

\textbf{Report recruitment source.} Where and how were participants recruited? Popular sources include dedicated online participant pools, such as Prolific~\cite{peer2021data} and Mechanical Turk~\cite{buhrmester2011amazon}. Given recent concerns about financial exploitation of online participants~\cite{semuels2018internet, hara2018data}, some recruitment platforms---including Prolific---incorporate guarantees for fair compensation rates. {Beyond these dedicated platforms, projects may choose to engage sources such as student populations at academic institutions~\cite{druckman2011students} or online communities and forums~\cite{bartneck2015comparing, shatz2017fast}.} Broader populations {constitute} the recruitment source for projects leveraging {representative} panel methods.

\textbf{Report sampling method.} What approach was used to sample participants? In combination with recruitment source, the sampling method of a study helps determine study generalizability. Certain sampling methods, including representative panel approaches, strengthen the generalizability of study results. Others, including convenience or snowball sampling, can reduce costs and facilitate recruitment at the expense of generalizability.

\textbf{Report study duration.} How long did participants take to complete the study, on average?

\textbf{Report compensation details.} What was the average compensation provided to participants? When appropriate, include details about base compensation, average bonus, and minimum hourly compensation.

\textbf{Report procedure and methods.} What steps did the study follow? Which materials did the study use (e.g., tasks or questions for participants)? These details closely relate to experimental replicability. Researchers should provide materials where possible (e.g., screenshots of participant interface) or refer to other sources that allow other researchers to re-implement the protocol.

\label{sec:guidelines_end}

\section{Conclusion} \label{sec:conclusion}

AI research with human participants currently falls short of the ethical standards expected of behavioral research. To help address this gap, this paper explores both normative concerns in AI research and {the history of lessons from other research fields} to construct a guiding framework for ethical AI research with human participants. This framework identifies four key principles (\textit{autonomy}, \textit{beneficence}, \textit{justice}, and \textit{accountability}) and a set of practical guidelines before, during, and after the study to help AI researchers navigate their work amidst the context of participatory design, dataset crowdsourcing, and the major role played by companies in research and development efforts.

The principles and guidelines described here are neither exhaustive nor finalized. They provide no formal guarantee that a research study provides a dignifying experience for participants, but should help foster ethical, responsible interactions between participants and researchers. Additions, modifications, and deletions will likely prove necessary over time.
For instance, given the influence of participatory design on AI research, \textit{co-partnership}---in the sense of community participation in research agenda-setting---may prove a promising addition. Do AI researchers have a responsibility to partner with research participants when identifying goals and ideal outcomes of research programs? A co-partnership principle could expand beyond the focus on protections encoded by existing principles and help advance the difficult work of aligning AI systems with human values~\cite{gabriel2020artificial}.

{This paper represents an initial foray into the complex ethics landscape of AI research with human participants, beginning with an exploratory analysis of current practices and culminating in a proposed ethics framework. Future work should seek to improve our understanding of current ethical norms---both by more rigorously investigating research papers and reports (e.g., pairing automated analysis with expert review), and by consulting researchers themselves (e.g., seeking insights through surveys and interviews). On the conceptual side, the framework presented here would benefit from additional refinement and contextualization. Future iterations should prioritize the integration of perspectives from a broad community of stakeholders, including AI researchers, ethicists, social scientists, and participants themselves.}

Overall, the framework presented here is a starting point for AI researchers to responsibly conduct research with human participants. Such research efforts are likely to increase in scope and pace as contemporary AI continues to advance and expand. This ethical framework is likewise a conversation in progress---more discussion will follow.

\section*{Acknowledgements}
I am indebted to Mark Díaz, Seliem El-Sayed, Sara McLoughlin Figel, Julia Haas, Will Hawkins, Arianna Manzini, Piotr Mirowski, Patrick Pilarski, and Laura Weidinger for their insightful comments and feedback on this work, and to Manuel Kroiss and Markus Kunesch for technical support on the paper annotation process.

\printbibliography

\begin{IEEEbiography}[{\includegraphics[width=1in,height=1.25in,clip,keepaspectratio]{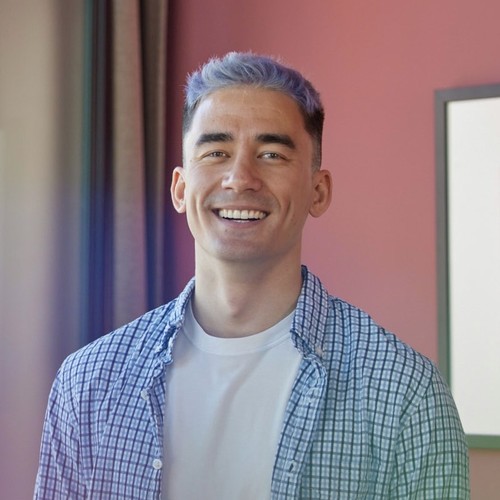}}]{Kevin R. McKee} received the A.B. degree in psychology from Princeton University, Princeton, NJ, USA, in 2014.

He currently works as a Staff Research Scientist at Google DeepMind, focusing in particular on developing cooperative and inclusive AI. His recent work explores the alignment of AI systems with human values, the ethical and epistemic risks of replacing human research participants with AI surrogates, and the development of sociotechnical methods to evaluate and align large language models.\end{IEEEbiography}

\begin{appendices}
\renewcommand{\thesubsection}{\thesection\arabic{subsection}}
\renewcommand\thesubsectiondis{\arabic{subsection}.}

\begin{figure*}[!b]
    \centering
    \includegraphics[width=0.8\textwidth]{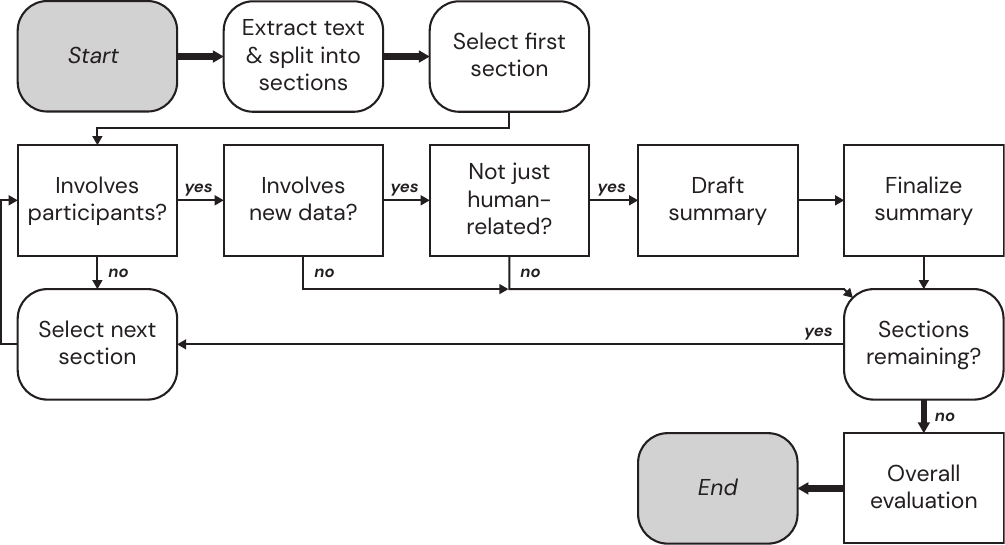}
    \caption{The automated annotation process applied to each paper. Boxes with sharp corners represent steps that produce annotations by applying custom prompts to the instruction-tuned large language model (LLM). Boxes with rounded corners depict steps that do not include the LLM. See Appendix~\ref{sec:human_data/prompts} for the exact text used in prompts.} \label{fig:annotation_flowchart}
\end{figure*}

\section{Evaluating current practices for human-participants research} \label{sec:human_data}

Proceedings from recent AI research conferences provide a valuable window into current practices in work with human participants. The present survey examined proceedings from the 2021, 2022, and 2023 AAAI conferences and the 2021 and 2022 NeurIPS conferences. The number of research papers in each of these proceedings is quite high (e.g., 1319 and 2671 papers at the 2022 AAAI and NeurIPS conferences, respectively). The papers are long-form (typically 6--8 pages of main text), with no guarantees that the involvement of human participants will be mentioned in the abstract.

Given these difficulties, \citeauthor{hawkins2023ethical}~\cite{hawkins2023ethical} applied an automated keyword analysis to annotate practices around human data collection at NeurIPS and other AI research venues. A keyword-based approach can efficiently process and annotate large number of papers, but is relatively inflexible: the initial selection of keywords may lead to large gaps in coverage (i.e., false negatives).

To attempt to iterate on this approach, the current evaluation automates the annotation process with an instruction-tuned large language model (LLM; see also~\cite{ouyang2022training}) and a chained annotation process (cf.~\cite{wei2022chain}). The evaluation pipeline iterated through each paper in the proceedings for the specified conferences (Figure~\ref{fig:annotation_flowchart}).

For each paper, the annotation process first extracted text from the paper~\cite{pdfium2023pdfium}. The next step partitioned the text into sections of 1000 characters (roughly 250 tokens for the LLM). Each section then passed through a conditional chain of prompts to check for the involvement of human participants. The assessments of each section were combined and fed back into the LLM for an overall evaluation for the paper. If this overall evaluation indicated that the paper involved human participants, the annotation process subsequently applied a secondary set of prompts to assess the reporting of independent ethical review, collection of informed consent, and participant compensation.

This evaluation processed {a total of} 9556 papers{: 4551} from the 2021, 2022, and 2023 AAAI {conferences,} and {5005 from the} 2021 and 2022 NeurIPS conferences{. Overall, 838 papers (8.8\%) involved the collection of original data from human participants.}
Table~\ref{tab:papers_summary}
{reports topline results concerning the proportions of papers reporting the implementation of independent ethical review, the collection of informed consent, and participant compensation}. Spot checks of the model explanations identified and removed a small number of false positives before the final tabulation.

In addition to the results reported in
Table~\ref{tab:papers_summary},
this investigation produced several new insights. Overall, the automated annotation process estimates a rate of human data collection that is somewhat higher than appraised by \citeauthor{hawkins2023ethical}. This increased estimate may stem from several different causes. First, the evaluation revealed several terms indicative of human participation beyond those identified by \citeauthor{hawkins2023ethical}, including ``volunteer'' and ``user study''. Second, the process of extracting text from conference papers is error-prone (e.g., the addition of an incorrect character or misplacement of a space). While minor transcription or spelling errors can disrupt keyword matching, the LLM proved generally robust to these errors. Third, while many papers mentioned human participation in their abstracts, a substantial number only disclosed this information within the methods or results sections---sometimes with just a single sentence. This final pattern underscores the risks of focusing solely on abstracts when assessing the involvement of human participants in research.

Overall, this exploratory analysis indicates that the recruitment of human participants for AI research is more prevalent than previously estimated, and confirms that ethical practices are not widespread in AI and ML research, despite the community's proximity to the fields of psychology and HCI.

\subsection{Prompts} \label{sec:human_data/prompts}

As summarized in Figure~\ref{fig:annotation_flowchart}, the annotation process applied an LLM and a series of prompts to evaluate the involvement of human participants in research papers. The following subsections lay out the prompts used in this process, with ellipses indicating a formatted input (e.g., text from the paper or from previous annotations). Because LLMs can generate complex text in response to these prompts, this process produces a number of auxiliary outputs in the form of the model explanations. These explanations can facilitate the process of quality review.

\subsection*{\textit{``Involves participants?'' prompt}}

Question: The following text is a section of a research paper. Does this section indicate that the research included data collection from human participants, annotators, user studies, etc.? This is often, though not always, revealed through key terms like ``participant'', ``annotator'', ``labeller'', ``user study'', or recruitment sources like ``Mechanical Turk''. Research involving ``subjective evaluation'' or ``perceived'' metrics may also rely on human participants to produce these measurements.

Text: ``...''

Reminder: Start your answer with ``this section does not involve participants, annotators, etc.'', or with ``this section involves'' followed by an explanation of exactly how you know the section involves human participants in less than 20 words. Ignore reference sections.

\subsection*{\textit{``Involves new data?'' prompt}}

Question: The following text is a section of a research paper. Does this section indicate that the research included original data collection from human participants, annotators, user studies, etc., or just the re-use of human data collected from other sources? Only say that this section includes original data collection if you are absolutely sure the data is original rather than re-used: do not include instances where other papers or citations collect the human data, or where the section merely mentions a new dataset or new annotations.

Text: ``...''

Previous notes: ``...''

Reminder: Start your answer with ``this section describes the re-use of data'', or with ``this section describes'' followed by an explanation of exactly how you know the section involves original data collection in less than 20 words.

\subsection*{\textit{``Not just human-related topic?'' prompt}}

Question: The following text is a section of a research paper. Does this section indicate that the research included the active involvement of human participants in data collection, or does it just discuss human-related topics?

Text: ``...''

Previous notes: ``...''

Reminder: Start your answer with ``this section just discusses human-related topics'', or with ``this section describes'' followed by an explanation of exactly how you know the section involves the active involvement of human participants in less than 20 words.

\subsection*{\textit{``Draft summary'' prompt}}

Question: The following notes describe a section of a research paper. Please combine the notes to answer the question: Does this section indicate that the research included data collection from human participants, annotators, user studies, etc.? Make sure to preserve details about whether the data collected is original or re-used, as well as whether the section actively involves human participants, annotators, user studies, etc. or just discusses human-related topics.

First note: ``...''

Second note: ``...''

Third note: ``...''

Reminder: Do not make up new information.

\subsection*{\textit{``Finalize summary'' prompt}}

Question: Given the following text, is this statement accurate or inaccurate? Make sure to preserve details about whether the data collected is original or re-used, as well as whether the section actively involves human participants, annotators, user studies, etc., or just discusses human-related topics.

Text: ``...''

Statement: ``...''

Reminder: Start your answer with ``Yes, this section describes'' and then explain in less than 20 words, or state ``this section does not involve participants, annotators, etc.''

\subsection*{\textit{``Overall evaluation'' prompt}}

Question: The following statements are all judgments of sections from a research paper. Each statement assesses a different section of the paper, indicating whether the paper includes work with or new data collection from human participants, annotators, labelers, user studies, etc. Based on these statements, do you think the paper actually includes new data collection from human participants, annotators, labelers, user studies, etc.? Ignore the re-use of human data collected from other sources and sections that just discuss human-related topics.

Statements: [...]

Options: [``yes'', ``no'']

Reminder: Ignore datasets that do not involve human participants, annotators, etc., even if they are new datasets. Answer only ``yes'' or ``no'', with no additional text or explanation.

\newpage
\section{Placing research ethics for human participants in historical context} \label{sec:what_is_the_historical_context}

What historical moments contributed to the particular ethical concerns highlighted by the modern research ethics movement? The events described in this appendix raised new, provocative concerns for scientists, in some cases prompting the formalization of ethical frameworks to reorient power relations between researchers and participants. These provocations---around informed consent, participant well-being, and related values---foreshadow many of the principles and guidelines outlined in this paper.

\subsection{Early incidents in biomedical research}

Two major incidents in the 20th century jump-started the modern research ethics movement: Nazi medical experiments during the Holocaust and the US Public Health Service Syphilis Study at Tuskegee. Revelations concerning these biomedical research projects raised new provocations around participant well-being and informed consent.

During World War Two, Nazi physicians forced large numbers of prisoners into medical experiments~\cite{gillespie1989research}. The procedures involved in the experiments were dangerous, abusive, and painful---causing extraordinary rates of death, disability, and disfiguration among the many victims. After the war ended, US military tribunals judged the physicians involved in the experiments at the war crime trials held in Nuremberg. The so-called doctors’ trial resulted in prison and death sentences for the majority of the physicians, though some were acquitted. The tribunals subsequently produced what is now known as the Nuremberg Code: a set of 10 points concerning ethical research conduct~\cite{nuremberg1947nuremberg}. Given the atrocities committed in the Nazi medical experiments, the Code argued for multiple standards protecting participant well-being, including the ``avoidance of unnecessary pain and suffering'' and the ``belief that [...] experimentation will not end in death or disability.'' The Nuremberg Code also enshrined a strong conviction that ``the voluntary consent of the human subject is absolutely essential.'' Though the concept of informed consent predates World War Two, the Nuremberg trials catalyzed the modern stance that informed consent should be the primary basis for ethical research with human participants~\cite{vollmann1996informed}.

\begin{figure}[t]
	\centering
    \includegraphics[width=0.4\textwidth]{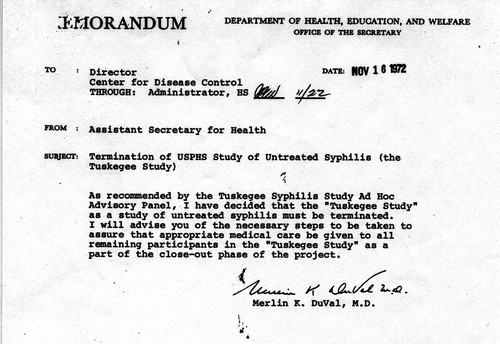}
	\caption{The letter from the Center for Disease Control ending the US Public Health Service study at Tuskegee. Source: US National Archives.}
	\label{fig:cdc_moratorium}
\end{figure}

In 1932, the US Public Health Service recruited several hundred Black American men from disadvantaged communities in Tuskegee for biomedical research on syphilis~\cite{brandt1978racism}. Officials promised the men free medical care for their involvement in the study, but did not inform them of their syphilis diagnoses. The study lasted four decades; throughout this time, the study researchers did not provide treatment to any of the infected men, even at the point penicillin was established as an effective (and readily available) treatment. The study ended in 1972 (see Figure~\ref{fig:cdc_moratorium}) after a social worker leaked details of the affair to the press~\cite{gamble1993legacy}. As with the experiments conducted during World War Two, revelation of the treatment of the men at Tuskegee prompted public outcry. Given the denial of available treatment to patients, the US Public Health Service study re-emphasized the importance of prioritizing participant well-being and the avoidance of unnecessary risks. The investigators’ exploitation of a disadvantaged community based on race was particularly troubling. As \citeauthor{brandt1978racism}~\cite{brandt1978racism} lambasted, the study ``revealed more about the pathology of racism than the pathology of syphilis.'' In 1974, as a result of the ethical lapses reflected in the investigators’ actions, the US Congress established the National Commission for the Protection of Human Subjects. Building on the framework laid by the Nuremberg Code, the Committee went on to issue a number of reports and recommendations to establish stronger standards for research ethics with human participants, most notably including the Belmont Report. 

\subsection{Mid-century dilemmas in psychology research}

\begin{figure}[!t]
	\centering
    \includegraphics[width=0.39\textwidth]{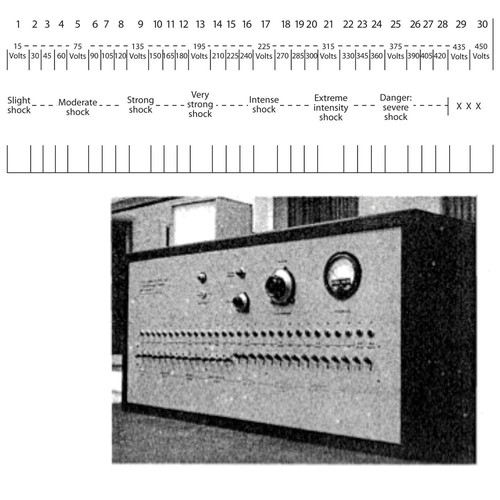}
	\caption{The ``shock machine'' used in Milgram's experiments. Source: \cite{saylor2012principles}.}
	\label{fig:milgram_stimuli}
\end{figure}

Following the revelations within the Nuremberg trials, social science found itself profoundly affected by the atrocities committed in the Holocaust. Many scientists took up new research programs to better understand how such events could come to pass. Stanley Milgram, an American psychologist, was one such researcher. Milgram ruminated on the widespread belief that ``associated the occurrence of the Holocaust with blind obedience to authority'' and stereotyped Germans as particularly conformist or obedient~\cite{russell2011milgram}. To evaluate this theory empirically, Milgram devised an experiment to measure an individual’s inclination to obey authority. In his experiment, an investigator told participants that they had been randomly assigned a ``teacher'' role and needed to administer questions to another participant, the ``learner.'' Every time the learner answered a question incorrectly, the learner was to give them an electric shock of gradually increasing voltage. Labels on the shock machine described the shocks, starting with ``Slight shock'' and ending with ``XXX'' (see Figure~\ref{fig:milgram_stimuli}). In actuality, the learner was an actor hired by Milgram and received no actual shocks. Nonetheless, his reactions were convincing. The investigator would respond to any participant inquiry or resistance by stating: ``I’m responsible for anything that happens to [the learner]. Continue please.'' Against his expectations, Milgram found that the majority of his participants would proceed all the way to administering the maximum shock. After conducting a series of experimental variations, Milgram published his results in three journal papers and most notably in his 1974 book \textit{Obedience to Authority}~\cite{milgram1974obedience}.

Though initially venerated in psychology, recent perspectives raise a number of questions about the ethics of Milgram’s experiments. The experiments purposefully led participants to believe that they were causing pain and suffering. As Milgram’s own notes admit, participants ``indicated that it caused them discomfort to watch the victim in agony''~\cite{russell2011milgram}. Others ``recalled the anxiety and distress they felt over an extended period at having to live with the knowledge that they may have killed someone''~\cite{brannigan2015introduction}. How justifiable was it for the study to cause this level of distress in participants? Milgram’s approach to deception compounds the issue. Milgram repeatedly insisted that he debriefed all participants in the experiments. However, subsequent archival review of the study’s records clearly demonstrate that Milgram did not debrief all of his participants and inform them that they had not harmed the actor, at least in part to ensure knowledge of the deception did not bias subsequent participants~\cite{nicholson2011torture}.\footnote{This conduct is particularly surprising given Milgram’s disavowal of deception during his graduate studies, writing that ``deception for any purposes [in psychology] is unethical and tends to weaken the fabric of confidence so desirable in human relations''~\cite{brannigan2015introduction}.}

\subsection{New questions on industry and oversight in the 21st century}

In 2014, a team of researchers from Facebook and Cornell University published an experimental paper entitled ``Experimental evidence of massive-scale emotional contagion through social networks'' in the \textit{Proceedings of the National Academy of Sciences}~\cite{kramer2014experimental}. In the paper, the researchers described controlled interventions to shift the emotional valence of Facebook users’ newsfeeds. The study aimed to understand whether emotional contagion occurs in online social networks. Soon after publication, the paper prompted considerable public criticism. Attention primarily coalesced around whether the users involved in the study had provided informed consent. The research team held that users’ agreement to Facebook’s data use policy constituted informed consent, as one term in the policy referred to the use of user information ``for internal operations, including troubleshooting, data analysis, testing, research and service improvement''~\cite{shaw2016facebook}. Critics argued that this approach diverged from ethical standards in academic research, which emphasizes transparent and comprehensible communication around study participation~\cite{flick2016informed}. Overall, the discussion drew attention to tensions at the intersection of psychology, research oversight, and industry versus academic standards for research ethics.

\subsection{Regulatory artifacts and current trends}

\begin{figure}[!t]
	\centering
    \includegraphics[width=0.2925\textwidth]{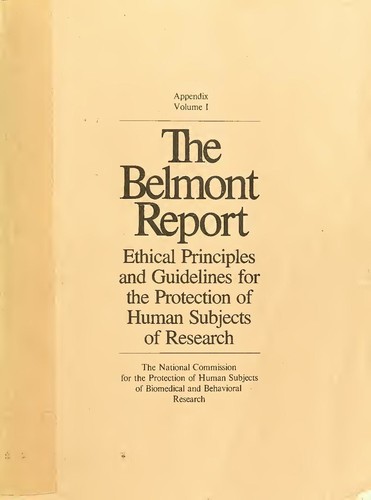}
	\caption{The cover of the Belmont Report. Source: The Internet Archive.}
	\label{fig:belmont_report}
\end{figure}

Together, these historical moments produced a series of advisory and regulatory artifacts that continue to steer modern research ethics. The Belmont Report (Figure~\ref{fig:belmont_report}) and the Common Rule, both developed in the United States, exert a particularly strong influence on current thought and practices---in the United States and beyond.\footnote{See also \citeauthor{holm2020belmont}~\cite{holm2020belmont}, who argues that the Belmont Report exerted a clear but indirect influence on the development of research ethics in Europe, but notes that European behavioral researchers were largely able to resist the formal establishment of regulatory frameworks, diverging from American behavioral scientists.} The framework introduced in the Belmont Report acted as the basis of the 1991 Federal Policy for the Protection of Human Subjects, more commonly known as the ``Common Rule.'' The three core ethical principles identified by the Belmont Report---respect for persons, beneficence, and justice---continue to organize federal regulation, persisting through the 2018 revision of the Common Rule to address new technologies and concerns. Formally, the Common Rule pertains only to research funded by the United States government. In practice, virtually all US academic institutions apply the Common Rule to research projects with human participants, irrespective of funding source. Similar principles and guidance appears in codes of ethical conduct in many other countries~\cite{shmueli2021beyond}.

\clearpage
\section{Defining the scope of research participation in AI~research} \label{sec:who_are_participants}

Humans enter many points of the process of designing, developing, and deploying AI systems. Research participants represent a particular subset of these entry points---that is, a particular subset of humans involved in research studies. Within the United States, the Department of Health and Human Services (HHS) drives standards and norms for research involving human participants, largely as laid out in the Federal Policy for the Protection of Human Subjects (the ``Common Rule''). These standards exert broad influence in the global setting; national regulations in many other countries are modeled after the Common Rule~\cite{capron2008legal}. Understanding the scope of AI research with human participants, then, can begin with the definitions offered by the Common Rule.

The Common Rule defines the scope of ``research'' as ``a systematic investigation, including research development, testing, and evaluation, designed to develop or contribute to generalizable knowledge'' (45 C.F.R. § 46.102). Based on this definition, the Common Rule specifies that a ``human subject'' is:

\begin{displayquote}
``a living individual about whom an investigator (whether professional or student) conducting research: (i) Obtains information or biospecimens through intervention or interaction with the individual, and uses, studies, or analyzes the information or biospecimens; or (ii) Obtains, uses, studies, analyzes, or generates identifiable private information or identifiable biospecimens.'' (45 C.F.R. § 46.102)
\end{displayquote}

As described in
Section~\ref{sec:contextual_concerns},
this paper adopts an adapted version of the Common Rule definition to scope human participation for AI research. Primarily, the adaptation shifts away from ``biospecimens'' and focuses on the general category of ``information'':

\begin{displayquote}
Participants are living individuals who provide or contribute information to researchers, particularly in systematic investigations designed to develop or contribute to generalizable knowledge.
\end{displayquote}

Teams at technology companies undertake a large proportion of contemporary AI research~\cite{birhane2022values, luitse2021great, martinez2021research}. The scoping of research to investigations intended to ``contribute to generalizable knowledge'' is sometimes used to excuse industry researchers from research ethics frameworks~\cite{jackman2015evolving, meyer2020there}, since they do not always intend technology and knowledge production for scientific dissemination~\cite{evans2010industry}. This ``non-generalizable'' defense typically casts industry projects as aiming to improve products (e.g., A/B testing) rather than for general knowledge, thus exempting them from research ethics frameworks. Two general patterns challenge this exclusion. First, there is growing recognition among ethicists that ``generalization'' is a limited diagnostic test to identify research. That is, generalization to broader reference groups is not the only objective in psychology and behavioral research. For example, research may also strive to generate a deeper understanding of the processes underlying a phenomenon or to test the sufficiency of certain conditions in bringing about an effect~\cite{mook1983defense}. Second, many of the technology companies and commercial organizations involved in AI research espouse goals that are more general than simply improving products, such as creating safe AI systems that benefit humanity~\cite{gdmndabout, openaindcharter}. In this sense, research conducted with human participants at these companies indeed aims to develop or contribute to generalizable knowledge. Thus, in several respects, the AI work undertaken in industry merits scrutiny within research-ethics frameworks.

On a qualitative level, research typically follows several procedural and methodological patterns. A common characteristic of research with human participants is submission and publication at peer-reviewed, scholarly venues, though releases in pre-prints and technical blogs are increasingly common. AI studies involving human participants also tend to follow established research designs, incorporating ``traditional'' psychology and human-computer interaction methods.

Many projects conduct behavioral, cognitive, or psychological experiments~\cite{campbell1963experimental, fisher1935design}. The hallmark of experimental studies is the controlled manipulation of a variable of interest---the independent variable. Experiments with human participants expose participants to particular values of the independent variable and measure changes in one or more dependent variables. Researchers design the experiment to minimize the influence all other environmental variables, typically by holding each at a constant value across all participants. For example, \citeauthor{strouse2021collaborating}~\cite{strouse2021collaborating} recruited a sample of $N = 114$ online participants for an experiment on human-agent collaboration. Their experiment randomly paired participants with various agents to test the hypothesis that certain training algorithms supported better human-agent collaboration.

Questionnaires~\cite{paulhus2007self} and surveys~\cite{visser2000survey} are also common. Questionnaires often feature in observational studies, though they can also be integrated within experimental studies. For example, \citeauthor{smith2021effects}~\cite{smith2021effects} investigated perceptions of autonomy and creativity in the context of generative music systems with a sample of $N = 5$. They selected their participants from a pool of Georgia Tech Music Technology graduate students. The primary task involved participants rehearsing a musical performance with neural network-based generative music systems. Participants answered questionnaires both before and after the music-performance task, allowing the researchers to understand participants’ background with machine learning concepts, as well as their reactions to the particular generative systems in the study. 

AI research with human participants may also draw on techniques such as semi-structured interviews~\cite{fontana2000interview} and focus groups~\cite{parker2006focus}. \citeauthor{gero2020mental}~\cite{gero2020mental} provide an example in this space, conducting semi-structured interviews to understand people’s mental models of AI agents. In one study, they recruited $N = 11$ participants from within a technology company. Participants first played a game with a cooperative AI agent, then completed interviews that explored their perceptions and understanding of the agent.

\clearpage
\section{A note on terminology} \label{sec:terminology}

Scientists have long debated how to refer to individuals participating in research~\cite{bibace2009name}, including the choice of ``subjects'' or ``participants.'' US federal regulation, including the Common Rule, adopts the term ``subjects'' (45 C.F.R. § 46.102). As a result, ``human subjects'' terminology is frequently seen in areas governed or influenced by the Common Rule---including the official guidance offered at many academic institutions.

Some scholars and stakeholders argue that ``subjects'' terminology carries negative implications concerning individual agency and autonomy. Using ``participants'' is thought to more directly recognize that people should participate in---not be subjected to---research. For example, the British Psychological Society advocates for the use of ``participant'' over ``subject'', as the former better ``serves to acknowledge the autonomy and agency of the individual in contributing to the research''~\cite{oates2021bps}. The American Psychological Association maintains a slightly more circumspect position, but advises the use of language that acknowledges the ``contributions and agency'' of research participants~\cite{apa2019participation}.
These tensions emerge in other linguistic norms, such as the passive language with which the Common Rule discusses participants, defining them as ``individuals about whom an investigator [...] obtains information'' (rather than ``individuals who contribute information'' or ``who produce information then collected by investigators'', for instance).
This approach reflects an atavistic connection to historical perspectives in psychology: both Wilhelm Wundt and Franz Brentano viewed their study participants as social equals who actively provided researchers ``the privilege of access into their innermost feelings and thoughts''~\cite{bibace2009name}.

Recognizing that the language of research itself can reinforce harmful power structures~\cite{morrison1994nobel}, this paper adopts ``participant'' terminology except in cases where directly quoting a source’s use of ``subject'' or ``subjects.''

\end{appendices}

\end{document}